# On the material time derivative of volume, surface, and line integrals

K.Y. Volokh

Faculty of Civil and Environmental Engineering, Technion-I.I.T., Haifa 32000, Israel

(E-mail: cvolokh@technion.ac.il )

**Abstract**

Traditional derivation of the material time derivative of volume, surface, and line integrals relies upon the notion of a referential configuration of continuum. Such a notion, however, is artificial and, probably, somewhat misleading in cases of liquids, gases, plastic flow of solids etc. It is, consequently, desirable to separate the formal calculation of the material time derivative of time-dependent integrals from any physics that can be related with them. Such a separation is targeted in the present letter where no referential continuum configuration or other physical notion is involved. *The material time derivative of volume, surface, and line integrals is presented as a purely mathematical manipulation*.

**1. Introduction**

Traditional derivation of the *material* time derivative of volume, surface, and line integrals *over a set of continuum particles* relies upon the notion of a referential configuration of continuum: Chadwick (1976); Eringen (1962); Gurtin (1981); Jaunzemis (1967); Liu (2002); Malvern (1969); Truesdell and Toupin (1961); Wilmanski (1998). Such a notion[1], however, is artificial and, probably, somewhat misleading in cases of liquids, gases, plastic flow of solids etc. It is desirable to separate the formal calculation of the material time derivative of time-dependent integrals from any physics[2] related with them. Such a separation is targeted in the present note where no referential continuum configuration or other physical notion is involved. The material time derivative of volume, surface, and line integrals is presented as a purely mathematical manipulation.

---

[1] Moreover, some authors mix the material time derivative of the integrals with the mass conservation law what can be confusing for less experienced readers.

[2] Fluid mechanics texts rely heavily upon physical intuition: Landau and Lifshitz (1987), Batchelor (2000), Sedov (1971).

Our intention is to derive the kinetics of volume, surface, and line integrals in the Eulerian description, which does not label continuum particles by using a referential configuration: where a particle comes from and where it goes does not matter. Accordingly, we consider the current positions[3] $\mathbf{x}(t)$ and velocities $\mathbf{v} = \dot{\mathbf{x}}$ of particles in space without identifying particles by their referential positions.

We are looking for the material time derivative of the time-dependent volume, surface, and line integrals of field quantity $\psi(t, \mathbf{x}(t))$. We will use the Cartesian components of tensors for the sake of simplicity and clarity.

## 2. Material time derivative of a volume integral

We start with the calculation for the volume integral emphasizing the notational difference between the full/material time derivative $d\psi/dt \equiv \dot{\psi}$ and the partial time derivative $\partial \psi/\partial t$ where $\mathbf{x}$ is fixed.

$$\begin{aligned}
\frac{d}{dt}\int_{\Omega(t)} \psi(t) dV(t) &= \int_{\Omega(t)} \frac{d}{dt}(\psi(t) dx_1(t) dx_2(t) dx_3(t)) \\
&= \int_{\Omega(t)} (\dot{\psi} dx_1 dx_2 dx_3 + \psi d\dot{x}_1 dx_2 dx_3 + \psi dx_1 d\dot{x}_2 dx_3 + \psi dx_1 dx_2 d\dot{x}_3) \\
&= \int_{\Omega(t)} (\dot{\psi} + \psi\frac{\partial v_1}{\partial x_1} + \psi\frac{\partial v_2}{\partial x_2} + \psi\frac{\partial v_3}{\partial x_3}) dx_1 dx_2 dx_3 \\
&= \int_{\Omega(t)} (\dot{\psi} + \psi \operatorname{div} \mathbf{v}) dV
\end{aligned} \qquad (1)$$

where the use has been made of $(dx_i(t))^\bullet = d\dot{x}_i(t) = dv_i(t)$.

The last integrand in (1) can be rewritten as follows

$$\dot{\psi} + \psi \operatorname{div} \mathbf{v} = \frac{\partial \psi}{\partial t} + \frac{\partial \psi}{\partial x_i}\frac{\partial x_i}{\partial t} + \psi\frac{\partial v_i}{\partial x_i} = \frac{\partial \psi}{\partial t} + \frac{\partial \psi}{\partial x_i}v_i + \psi\frac{\partial v_i}{\partial x_i} = \frac{\partial \psi}{\partial t} + \frac{\partial(\psi v_i)}{\partial x_i}. \qquad (2)$$

Substituting (2) in (1) we have

$$\frac{d}{dt}\int_{\Omega(t)} \psi dV = \int_{\Omega(t)} \{\frac{\partial \psi}{\partial t} + \operatorname{div}(\psi \mathbf{v})\} dV . \qquad (3)$$

*Concerning the differentiation in (1) we remind that the integral is an infinite sum of the volume increments weighted by the integrand function and we can differentiate through the integral if the increment is included in the derivative as it is done after the first equality in (1).*

---

[3] We do not make a distinction between $\mathbf{x}$ as a position of continuum particle and a fixed space position. This distinction is always clear from the context.

## 3. Material time derivative of a surface integral

The calculation of the material time derivative of a surface integral is a more subtle matter than the calculation considered above. Our key idea is to use the divergence theorem to turn the surface integral into the volume integral at the beginning of the manipulation and to return from the volume integral to the surface integral at the end of the manipulation as follows

$$\frac{d}{dt}\int_{\partial\Omega(t)}\psi n_i dA = \frac{d}{dt}\int_{\Omega(t)}\frac{\partial\psi}{\partial x_i}dV$$

$$= \int_{\Omega(t)}\left\{\frac{\partial}{\partial t}\left(\frac{\partial\psi}{\partial x_i}\right)+\frac{\partial}{\partial x_j}\left(\frac{\partial\psi}{\partial x_i}v_j\right)\right\}dV$$

$$= \int_{\Omega(t)}\left\{\frac{\partial}{\partial x_i}\left(\frac{\partial\psi}{\partial t}\right)+\frac{\partial}{\partial x_j}\left(\frac{\partial\psi}{\partial x_i}v_j\right)\right\}dV$$

$$= \int_{\Omega(t)}\left\{\frac{\partial}{\partial x_i}\left(\dot{\psi}-\frac{\partial\psi}{\partial x_j}v_j\right)+\frac{\partial}{\partial x_j}\left(\frac{\partial\psi}{\partial x_i}v_j\right)\right\}dV$$

$$= \int_{\Omega(t)}\left\{\frac{\partial\dot{\psi}}{\partial x_i}-\frac{\partial}{\partial x_i}\left(\frac{\partial\psi}{\partial x_j}v_j\right)+\frac{\partial}{\partial x_j}\left(\frac{\partial\psi}{\partial x_i}v_j\right)\right\}dV$$

$$= \int_{\Omega(t)}\left\{\frac{\partial\dot{\psi}}{\partial x_i}+\frac{\partial}{\partial x_i}\left(\psi\frac{\partial v_j}{\partial x_j}\right)-\frac{\partial}{\partial x_j}\left(\psi\frac{\partial v_j}{\partial x_i}\right)\right\}dV$$

$$= \int_{\partial\Omega(t)}\left\{\dot{\psi}n_i+\psi\frac{\partial v_j}{\partial x_j}n_i-\psi\frac{\partial v_j}{\partial x_i}n_j\right\}dA \quad , \tag{4}$$

where (3) has been used in the second equality.

We can rewrite (4) in the coordinate-free form

$$\frac{d}{dt}\int_{\partial\Omega(t)}\psi\mathbf{n}dA = \int_{\partial\Omega(t)}(\dot{\psi}\mathbf{n}-\psi(\partial\mathbf{v}/\partial\mathbf{x})^T\mathbf{n}+\psi\mathbf{n}\,\text{div}\,\mathbf{v})dA. \tag{5}$$

## 4. Material time derivative of a line integral

This calculation is trivial

$$\frac{d}{dt}\int_{\Gamma(t)}\psi dx_i = \int_{\Gamma(t)}\frac{d}{dt}(\psi dx_i)$$

$$= \int_{\Gamma(t)}(\dot{\psi}dx_i+\psi d\dot{x}_i) \quad , \tag{6}$$

$$= \int_{\Gamma(t)}(\dot{\psi}dx_i+\psi\frac{\partial v_i}{\partial x_j}dx_j)$$

or

$$\frac{d}{dt}\int_{\Gamma(t)}\psi\,d\mathbf{x} = \int_{\Gamma(t)}(\dot{\psi}\,d\mathbf{x} + \psi\frac{\partial\mathbf{v}}{\partial\mathbf{x}}d\mathbf{x}). \tag{7}$$

**5. Conclusion**

We showed how to calculate the material time derivative of the volume, surface, and line integrals without noting the reference configuration of continuum or appealing to any kind of physical intuition.